\documentclass[a4paper,12pt]{extarticle}
\usepackage[margin=3cm]{geometry}
\usepackage{newtxtext}
\usepackage[slantedGreek]{newtxmath}
\usepackage[pdftex]{graphicx}
\usepackage{natbib}
\usepackage{url}
\DeclareUrlCommand\doi{\urlstyle{rm}}
\usepackage{donaldpapers}

\title{Donald Lynden-Bell CBE\\
  5th April 1935 -- 6th February 2018\\
Elected FRS 1978}

\author{Neil~Wyn~Evans\\
\\
\emph{\small Institute of Astronomy, University of Cambridge, Madingley Road, Cambridge CB3 0HA, UK}\\
}

\date{30 March 2020}
\begin{document}
\maketitle

\begin{abstract}
Donald Lynden-Bell's many contributions to astrophysics encompass general relativity, galactic dynamics, telescope design and observational astronomy. In the 1960s, his papers on stellar dynamics led to fundamental insights into the equilibria of elliptical galaxies, the growth of spiral patterns in disc galaxies and the stability of differentially rotating, self-gravitating flows. Donald introduced the ideas of \lq violent relaxation' and \lq the gravothermal catastrophe' in pioneering work on the thermodynamics of galaxies and negative heat capacities. He shared the inaugural Kavli Prize in Astrophysics in 2008 for his contributions to our understanding of quasars. His prediction that dead quasars or supermassive black holes may reside in the nuclei of nearby galaxies has been confirmed by multiple pieces of independent evidence. His work on accretion discs led to new insights into their workings, as well as the realisation that the infrared excess in T Tauri stars was caused by protostellar discs around these young stars.  He introduced the influential idea of monolithic collapse of a gas cloud as a formation mechanism for the Milky Way Galaxy. As this gave way to modern ideas of merging and accretion as drivers of galaxy formation, Donald was the first to realise the importance of tidal streams as measures of the past history and present day gravity field of the Galaxy. Though primarily a theorist, Donald participated in one of the first observational programs to measure the large-scale streaming of nearby galaxies. This led to the discovery of the \lq Great Attractor'. The depth and versatility of his contributions mark Donald out as one of the most influential and pre-eminent astronomers of his day.
\end{abstract}

\section{Upbringing and Education}

Donald Lynden-Bell was born on 5th April 1935 in Dover and christened in the Castle's chapel, one of the most ancient churches in England. His father, Lieutenant Colonel Lachlan Arthur Lynden-Bell, fought on the Western Front and in the Middle East during World War I and received a Military Cross. Donald's very early childhood was spent in Fort George, near Inverness, where his father was in command. It was a peripatetic life, as the family moved wherever his father was posted. Spells in Glasgow, Nairn and Broughty Ferry followed, before the outbreak of the Second World War. Despite being a professional soldier for 25 years, Donald's father was now too poorly sighted for active service, and so he was despatched around the country to plan coastal defences. Donald spent the War years at his grandparents' house in Stonebarrow Hill, east of Charmouth in Dorset. The strongest influences on his early life came from his mother, Monica Rose, and his elder sister, Jean. As a boy, he loved to collect fossils, shells, pressed flowers, moths and butterflies, supplementing his earlier interests in mechanical toys. Donald found reading difficult and did not share his mother's love of literature and poetry.  He finally managed to read at age nine, when his mother plotted a graph of number of words read each night and would not let him stop until the graph was high enough. Donald remained a slow and reluctant reader throughout his life, but he always loved graphs!

In 1944, Donald's father joined the John Lewis Partnership and settled in Thames Ditton. After a couple of terms at Down House, a prep school in Esher, Donald was sent to a boarding school in Somerset to avoid the dangers of the doodlebugs. Once he had overcome homesickness, Donald blossomed, enjoying early morning horse riding on the Quantocks. He became very interested in how things work, and took mechanical things to pieces, building models with his Meccano. Donald also found out that he was very good at mathematics, but his poor reading skills held him back in many other subjects. 

Nonetheless, he passed Common Entrance and went to Marlborough College in 1948. In those days, it was safe to bicycle on main roads, so it was possible to visit places up to 10 miles away on a weekend afternoon. Donald got to know the flowers and butterflies of the sweeping escarpment from Milk Hill down to Pewsey Vale, as well as the grand neolithic monuments of Avebury and Silbury Hill. Under the Labour Government, the School Certificate was replaced with O Levels, which at first could not be attempted until over the age of 16. As a result, Marlborough College decided that Donald's year should be accelerated to take the last School Certificate at age 14. This was a great blessing, as it allowed Donald to specialise in mathematics and science a year earlier, and give up subjects that he found uncongenial. At that time, Marlborough had some wonderful teachers in mathematics, including A.R.D. Ramsey (co-author of textbooks on mechanics) and E.G.H. Kempson (an Everest mountaineer who also taught climbing).  Both had profound influence on Donald. Ramsey convinced his pupils that the gift of mathematics was very special. His pupils should regard the great mathematicians of the past as people to emulate and surpass. The best mathematicians were an elite, and it was their moral duty to make the most of their gifts. Kempson taught Donald not just mathematics at a high level, but also rock climbing in Wales at the Idwal Slabs and the Glyderau. At the end of his time at Marlborough College, Kempson leant Donald a copy of James Hopwood Jeans' book, {\it Problems of Cosmogony and Stellar Dynamics}. It was a premonitory choice, as Donald's thesis work was to make startling contributions to this subject. A contemporary of Donald's, Terry Wall FRS 1969, recalls that Donald was already devoted to astronomy and very active in the school's Science Society.

\section{Undergraduate Days (1953-1957)}

Donald came up to Clare College, Cambridge in 1953. He convinced his Director of Studies, R.J. Eden, that he should be allowed to skip Part I of the Mathematical Tripos, most of which he had already covered in school. Donald joined the Mountaineering Club (CUMC), the Astronomical Society, the Archimedeans and the Trinity Mathematical Society, which at that time allowed Clare mathematicians to join. CUMC arranged climbing meets in Wales, the Lake District, Ben Nevis and the Alps. Donald felt a wonderful sense of companionship with those on the same rope, as well as a love of the freedom and serenity of the mountains. 

In those days, Heffers bookshop had a secondhand section on the top floor of their shop in the old Petty Cury. There, Donald bought copies of the books of Arthur Stanley Eddington FRS 1914, including {\it Space Time and Gravitation}, {\it The Mathematical Theory of Relativity} and {\it The Internal Constitution of the Stars}. He was attracted to theoretical astrophysics as mathematically challenging and physically fascinating. Donald's Ph.~D. thesis would later be dedicated to Eddington and his inspirational writing. Donald also found the quantum theory intriguing, but the interpretation of the mathematics deeply unsatisfying. It gave up the idea of following what was happening in detail and substituted merely a calculation of the probabilities of the possible outcomes of an experiment. To him, it seemed the reverse of the disassembling of mechanical things to see how they worked.
  
Donald followed an eclectic choice of courses in his first year. Discovering that Hermann Bondi FRS 1959 was giving a Part III course on Relativity, he eagerly listened in. Though a freshman, he found Bondi's graduate-level lectures understandable, as they were physically based and beautifully delivered. Donald also discovered that lectures on Observational Astronomy were held at the Observatories in the afternoon. So, he attended well-delivered and authoritative lectures by Donald Blackwell on the intricacies of phenomenon observed on the solar surface. Perhaps because Donald was characteristically following his own path, he only achieved an upper Second at the end of the first year.

In his second year, Donald was supervised by two men who become very famous, namely Sir Michael Atiyah PRS 1990 and Abdus Salam (1979 Nobel prize winner in physics). Salam held the strong belief that one could only really learn physics from direct contact with those carrying out experiments, and firmly advised Donald to leave the Mathematical Tripos and switch to Physics Part II in the Natural Sciences Tripos for his third year. Brian Pippard FRS 1956, who was Director of Studies in Physics at Clare College, tried to persuade Donald to take two years over Part II, but Donald insisted on doing it in a year. He attended stimulating lectures by Otto Frisch FRS 1948 on nuclear theory, Martin Ryle FRS 1952 on Radio Astronomy and Nevill Mott FRS 1936 on quantum mechanics.  He learnt much from Joe Vinen FRS 1973, who had just performed his famous experiments on the quantised vortices of liquid helium. At the end of the examinations, Donald went climbing on Ben Nevis, ascending the difficult route over the mighty cliff below the summit. There was little communication with the outside world, so when he thought the results would be out, he descended to Fort William and bought a newspaper. He discovered that he had an upper second again, and so was unsure whether he could continue to research. Back in Cambridge, he was offered Ph.~D. positions to work on experimental studies of turbulence or on radioastronomy, but he eschewed both as too experimental for his tastes. He preferred the idea of taking Mathematics Part III with a view to doing a Ph.~D. in Theoretical Physics.

Although Donald's Part III year was full of interesting courses, he
did not have the same freedom of spirit to take lectures on what he
found most interesting. He had to learn the main courses really well
for the examinations -- Fred Hoyle FRS 1957 on Cosmology, Leon Mestel FRS 1977 on Cosmical Electrodynamics, Paul Dirac FRS 1930 on Quantum Field Theory and Nevill Mott on Atomic Collisions. This blunted his joy in finding things out. In the end, he obtained the desired distinction, but it was drudgery rather than love. He then found that most of the interesting potential supervisors in his preferred subject of quantum theory were about to go on leave. So, Donald decided to turn his long-standing hobby of astronomy into his work, by becoming a graduate student of Mestel, investigating the role of magnetism in astronomy

During his Part III year, Donald attended meetings of the Cambridge University Natural Sciences Club (CUNSC). This was limited to 13 people -- the maximum that would fit into an undergraduate room. Members read papers to one another about anything they found interesting, with the proviso that the lecture and discussion must be in terms everyone could understand. Donald found it interesting to learn more of the life sciences, especially botany and evolutionary biology. Through the club, he met Ruth Truscott FRS 2006, who he found full of energy and fun. He was also greatly delighted to find someone who \lq\lq enjoyed mathematics rather than feared it". Donald and Ruth were subsequently married four years later on 1 July 1961 in Mosely Parish Church, and remained great supporters of CUNSC which played an important part in both their lives.

\section{Ph. D. Days (1957-1960)}

While Donald learnt a great deal from Mestel, his thesis work did not
go well. His problem was to understand the effects of magnetic fields
on star formation, but neither Donald nor Mestel were familiar with
the necessary computational techniques to solve the awkward
integro-differential equation that confronted them. Donald commented
later that his \lq\lq algorithm was numerically unstable, the problem was physically unstable and the whole exercise not of prime importance for star formation''. After a frustrating year, Donald attended a summer course at the Royal Greenwich Observatory, then based at Herstmonceux Castle in Sussex. He struck up a friendship with the Astronomer Royal, Sir Richard Woolley FRS 1953, who set him some problems in stellar dynamics. Woolley was a good scientist and a natural leader of men. He was very encouraging when Donald made some progress. When nearly a year later, the work with Mestel had completely run into the sand, Donald already had an embryo paper based on Woolley's problem. This encouraged Donald to shelve the work with Mestel completely, and concentrate on new problems in stellar and galactic dynamics. This proved to be Donald's salvation. Some nine months later, in an inspired burst of creativity, Donald had solved some important and useful problems in stellar dynamics. He was able to present a research thesis for the Junior Research Fellowship competition at Clare College, which he won. Soon afterwards, he was also awarded a Harkness Fellowship. It was a remarkable nine months turnaround.

Donald's 1960 thesis is entitled {\it Stellar and Galactic Dynamics}. No first-rate mathematician had really looked at this subject since the days of James Jeans FRS 1906 and Eddington in the 1930s, so Donald was able to make substantial progress with analytical methods, always his forte. Jeans (anticipated by Poincar\'e) had already found that the distribution function of a collisionless stellar system must be a function of the integrals of motion only~\citep{BT}. Donald developed a new method to identify potentials admitting integrals of motion, and used it both to find new integrable systems and to categorise the ones already known~(\ref{LB1962a}). He laid emphasis on the importance of stellar systems with potentials separable in confocal ellipsoidal coordinates, first introduced into astronomy in an informal way by Eddington. Donald called these systems Eddington potentials, but they are now known in galactic dynamics texts as St\"ackel potentials~\citep{BT} -- "mistakenly \dots as he never derived them"~(\ref{LB2016}). Donald discovered many of the properties of stellar systems with potentials separable in ellipsoidal coordinates. Years later in the 1980s, these potentials were to attain a new vogue when Tim de Zeeuw, discovered the \lq perfect ellipsoid' on a working visit to Donald in Cambridge~(\ref{ZL85}). This is an exact model whose density law is similar to elliptical galaxies and whose potential is separable in ellipsoidal coordinates~\citep{Ze85}. The work retains relevance to this day, as modern stellar dynamics lays heavy emphasis on action-angle coordinates~\citep{Sa16}. Fast numerical algorithms have been developed to approximate the potential by a nearby St\"ackel potential and hence obtain actions and angles, and these form the basis of modern galactic dynamics software packages.

Donald also attacked the self-consistent problem of stellar dynamics -- that is, the finding of the phase space distribution function of a stellar system (which can depend only on the integrals of motion through the Jeans Theorem) consistent with the potential and density. This necessitates the solution of a Volterra integral equation of the first kind. If the stellar system is spherical, the distribution function can be extracted via an Abel transform, as Eddington had discovered in 1916. Donald showed how to find distribution functions for axisymmetric systems -- a much harder problem -- using Laplace transforms~(\ref{LB1962b}). He provided a flattened Plummer model (suggested by Maarten Schmidt) for which the calculations are all analytic and so created the first axisymmetric, self-consistent galaxy model with a distribution function. These ideas were later developed and extended by others~\citep{HQ,EV94}. Although Woolley deserves the credit for inspiring and encouraging Donald, it is clear that Donald's mind had already been prepared for these speedy discoveries in galactic dynamics by his wide reading of the books of Jeans and Eddington. As his nominal supervisor Mestel remarked years later to the author, \lq\lq Donald did not really need supervising".

\section{Harkness Fellow, CalTech (1960-1962)}

Donald took his Harkness Fellowship to CalTech. There he found congenial company with two other English postdocs, Wallace Sargent FRS 1981 and Roger Griffin. Together with Neville Woolf, the four organised a number of great camping and hiking trips in the American Southwest. This included trips to Death Valley, Meteor Crater and Rainbow Bridge. A famous photograph (reproduced in Donald's own biographical memoir of Wal Sargent) shows the four of them, joined by John Hazlehurst, having Christmas dinner in 1960 at the bottom of the Meteor crater, Arizona, surrounded by a Union Jack. In April 2011, the four men -- now grey with age -- reconvened on the 50th anniversary of the original expedition to Rainbow Bridge. The reunion was the basis of the film {\it Star Men} by Alison Rose, a graceful tribute to scientific passion, collegiality and ageing.

\begin{figure}
\includegraphics[width=\textwidth]{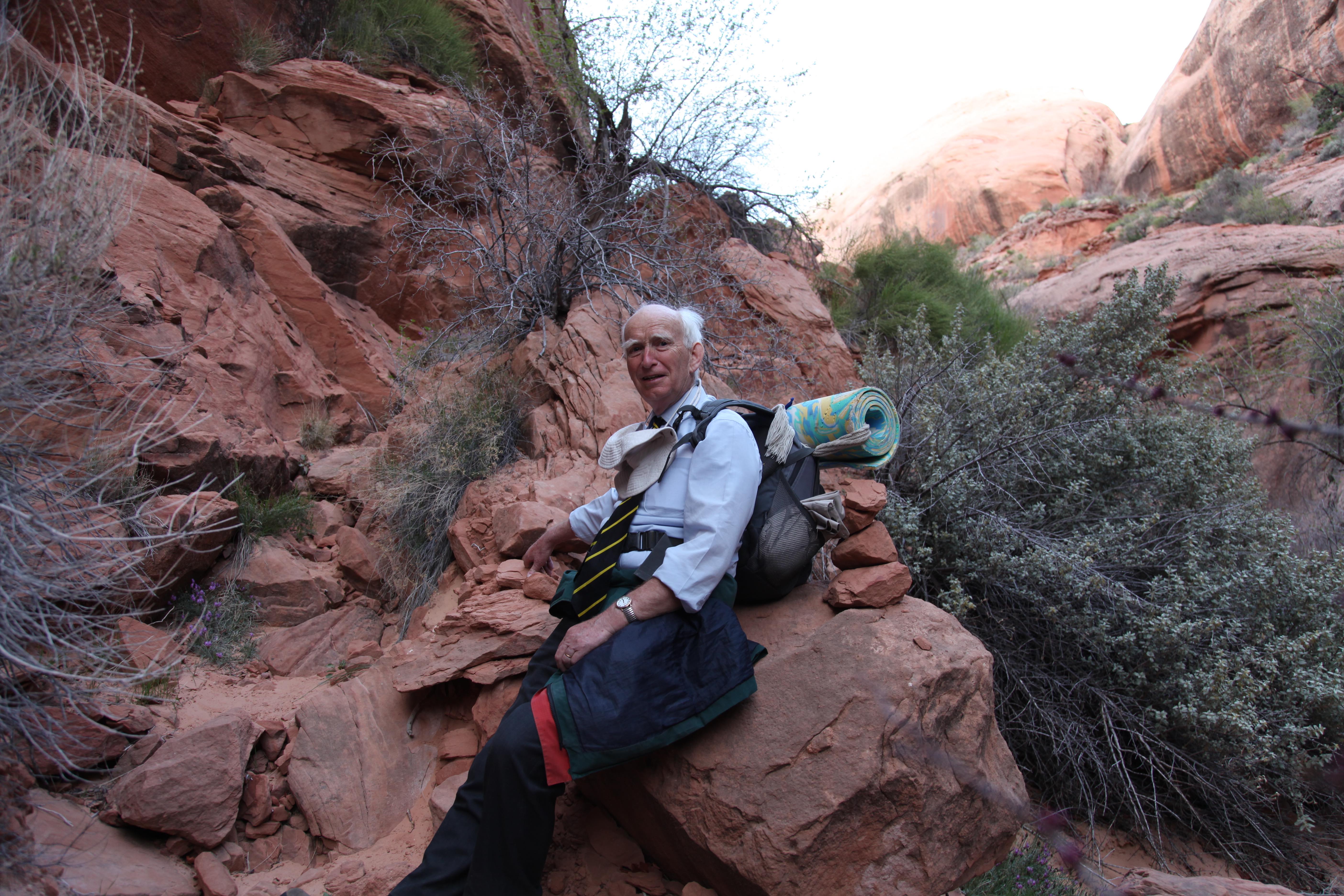}
\caption{Donald Lynden-Bell pictured in 2011 near Rainbow Bridge, Utah during the filming of Star Men, directed by Alison Rose. This recreated an adventure Donald had undertaken with Wal Sergent, Nic Woolf and Roger Griffin as postdocs in the 1960s. (Photo Credit: Inigo Films)}
\label{fig:staermen}
\end{figure}

In Caltech, Donald wrote one of his most famous papers~(\ref{ELS}), together with Olin Eggen and Allan Sandage and often referred to as ELS. They saw a strong correlation between the ultraviolet excess and the orbital eccentricity for a sample of high velocity stars in the solar neighbourhood. Stars with large ultraviolet excess have low metallicities. ELS found that -- locally -- the most metal-poor stars have the largest orbital eccentricities, the highest velocity dispersions and the lowest systemic rotation. In a provoking hypothesis, ELS suggested that the metal-deficient stars formed first in an approximately spherical configuration. This original gas cloud then collapsed quickly, giving birth to subsequent generations of stars. The contraction of the Galaxy not only yielded a population of younger stars on nearly circular orbits, but also made the orbits of the halo stars highly eccentric, thus neatly explaining the chemo-kinematical properties of their local stellar sample. This became the standard picture of galaxy formation in the 1960s and 1970s.

However, two aspects of the ELS picture became troubling. First, the timescale for collapse was estimated by ELS as the free-fall time or $\approx 2 \times 10^8$ years. This became difficult to reconcile with age spreads of Gigayears among the halo stars, suggesting a much more protracted formation. Second, the progressive chemical enrichment of the halo envisaged by ELS implied a strong correlation between metallicity and eccentricity for all halo stars, as well as a metallicity gradient with distance. Subsequent surveys, coupled with more precise determinations of stellar ages and elemental abundances, have revealed the chemo-dynamical history of the Galaxy in much greater detail than was available to ELS~\citep{SZ,Be95}. There is only a weak relationship between metallicity and eccentricity for halo stars, and there is no detectable metallicity gradient. It is now believed that the correlation between ultraviolet excess and eccentricity seen by ELS in the local sample may have been caused by a selection bias affecting the data~\citep{No85}. Notwithstanding this, the theoretical insights into the young Galaxy propounded by ELS remain as shrewd today as they were over 50 years ago.

The ELS picture of the formation of the Milky Way galaxy has been superseded by scenarios in which structure is accumulated hierarchically from smaller fragments. Alar \citet{To77} was already speculating~\lq\lq It seems almost inconceivable that there wasn't a great deal of merging of sizeable bits and pieces (including quite a few lesser galaxies) early in the career of every major galaxy no matter what it now looks like. The process would obviously have yielded halos from the stars already born, whereas any leftover gas would have settled quickly into new discs embedded within such piles of stars". If so, then the properties of the halo stars depend on their progenitor satellites, and no tight correlation between dynamics and chemistry is expected. Recent discoveries with astrometric data from the {\it Gaia} satellite provide irrefutable corroboration with the identification of the early mergers that built the Milky Way Galaxy~\citep{Be18, My19}. Although no longer viable as a mechanism, ELS has proven to be very influential. It was the first paper to emphasise that halo stars take very long times to exchange energy and angular momentum with the rest of the Galaxy. Hence, they retain memory of early times and so can be used to study the process of galaxy formation. It was also the first paper on the formation of the Galaxy that provided testable and quantitative predictions, and so it played its pioneering role in opening up a new discipline to scientific scrutiny.

As part of his contribution to ELS, Donald also developed a supremely elegant galaxy model, as recounted in (\ref{Ev90}). Unbeknowst to him, it had been independently found a few years earlier by Michel \citet{He59} and named~\lq L'Amas Isochrone'. The isochrone is the most general spherical potential for which the orbits, actions and angles of stars are calculable analytically as closed form expressions~(\ref{LBecc}). It is so named because the period of any star depends only on its energy, and not on its angular momentum. The model is still widely used as a simple, reasonably realistic representation of a galaxy and earns a rightful place in the standard texts on galactic dynamics~\citep{BT}.

\section{Assistant Lecturer, University of Cambridge (1962-1965)}

Donald returned to Cambridge in 1962 as a University Assistant Lecturer at the Department of Applied Mathematics and Theoretical Physics (DAMTP), as well as Director of Studies in Mathematics at Clare College. In his Ph.~D. thesis, he had early ideas on regenerative theories of spiral structure. He returned to this problem, with an engaging collaborator, Peter Goldreich, then a young NSF postdoctoral fellow in Cambridge. Together, they wrote two important papers (hereafter GLB) on the stability of gaseous disks, in the process discovering \lq\lq swing amplification''~(\ref{GLB1},\ref{GLB2}). Since the Earl of Rosse's discovery in 1851 of spiral patterns in the nearby disc galaxy M51, the persistence of spiral structure against differential rotation was a major unsolved problem. By the 1960s, the riddle of the spirals was attracting the attention of an impressive range of applied mathematicians, astrophysicists and fluid dynamicists, including C.C. Lin, Frank Shu, Chris Hunter, Agris Kalnajs and Alar Toomre. A swashbuckling recent history gives a flavour of those piratical times in spiral structure theory~\citep{Pa04}.

GLB's work on gaseous discs, together with the almost contemporaneous study on stellar discs by \citet{Ju66}, were major pieces in solving the puzzle. GLB found the unlocking key in a new shearing coordinate system, set up in a local patch of a differentially rotating disc. The co-moving axes are oriented radially and tangentially along the shearing flow, and the wave harmonics are sought in these new coordinates as solutions of a neat second-order ordinary differential equation. A leading spiral wave unwinds into a trailing wave thanks to the differential rotation. At early times, the intercrest spacing is small and gas pressure ensures stability. As the waves are swept round, the pressure loses its potency, the wavelength rises and the shear takes over. This amplifies the unwrapping wave into a vigorously growing, trailing spiral. The conspiracy between self-gravity, shearing due to differential rotation and gas pressure (or epicyclic motion for stars) was later christened swing amplification, thanks to a hat-tipping article by Alar~\citet{To81}, who first demonstrated the stellar dynamical analogues. In this picture, spirality arises from regenerative \lq\lq sheared gravitational instabilities". Spiral arms in galaxies are beautiful transients that bloom and fade. Tidal forcing by companions (such as NGC 5194 in the case of the Earl of Rosse's M51) or gravitational wakes around massive objects (like giant molecular clouds) provide the seeds. In fact, GLB's machinery was powerful enough to handle most of the later work of the Lin-Shu school, where density-wave theory was worked out using different methods~\citep{GT}.

Donald resumed collaboration with Goldreich some years later on a very different topic -- the then recently discovered bursts of decametric radiation from Jupiter, which are correlated with the phase of Io, its closest moon~(\ref{GLB3}). Jupiter's magnetic field moves rapidly past Io as it orbits. Io is a good conductor, so a current is produced by its motion through the Jovian magnetic field.  The plasma enclosed by the satellite's flux tube moves as though it were rigidly attached to Io. The electromotive force developed across its diameter due to its motion is transmitted along the flux tube which passes through the satellite and drives a current across each foot of the tube in the ionosphere. In Goldreich \& Lynden-Bell's model, it is the motion of the current up and down the flux tube that generates the decametric bursts through beam instabilities. This model broke new ground in showing how Io may play the role of a unipolar conductor and so control the bursts. The hot spot where Io's flux tube meets Jupiter was finally found. It pointed $15^\circ$ ahead of the orbital position of Io, just as they had predicted.

\begin{figure}
\includegraphics[width=\textwidth]{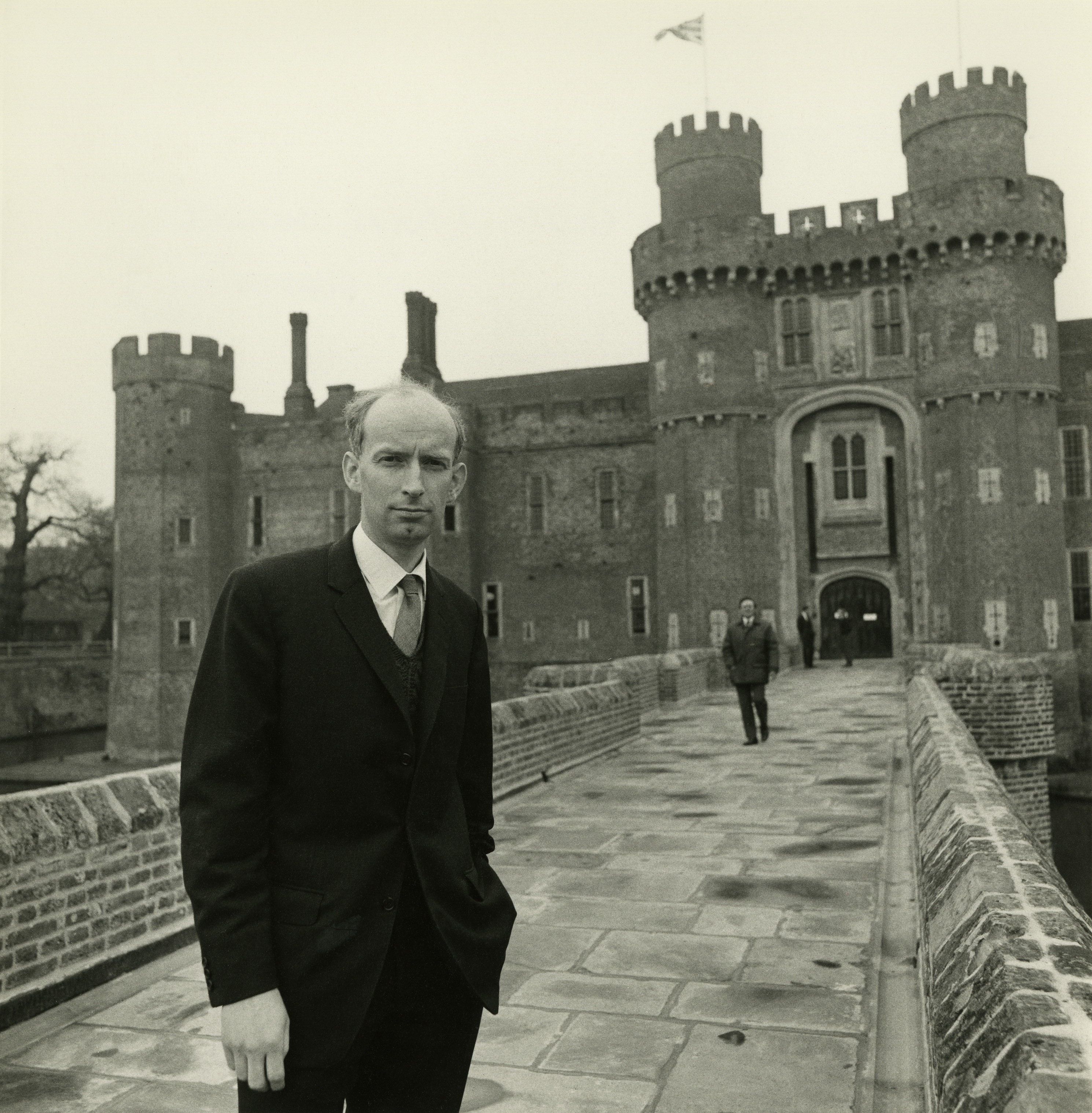}
\caption{Donald Lynden-Bell pictured at Herstmonceux Castle in the mid 1960s (Courtesy of Ruth Lynden-Bell)}
\label{fig:herst}
\end{figure}

\section{Royal Greenwich Observatory (1965-1972)}

As Donald recounted ruefully later,~\lq\lq it was not very long before
I discovered that three half-jobs, lecturing, researching and teaching for one's college, added up to one and a half rather than one" (\ref{LB2010}). To get more time for research, Donald decided to leave Cambridge and work under Woolley at the Royal Greenwich Observatory (RGO), then in Herstmonceux Castle in Sussex. No doubt Donald was also influenced by the inspirational role that Woolley played during his PhD years. His wife Ruth was appointed to a half-time lectureship in Chemistry at the new University of Sussex. They welcomed their daughter Marion to the family in 1965, and son Edward in 1968. Donald's research blossomed at the RGO in an intensely creative period under the stimulus of daily contact with observers in a working observatory.

\subsection{Violent Relaxation and Elliptical Galaxies}

Throughout the 1950s and 1960s, astronomers had been puzzled by the smooth light distributions of elliptical galaxies, given that the timescale for two-body relaxation exceeded their ages. This paradox is forcefully articulated in some of the textbooks of the time, such as {\it Dynamics of Stellar Systems}~\citep{Og65}. Donald solved this problem by showing that collective relaxation effects can result in a very rapid or \lq\lq violent relaxation"~(\ref{LB1967}). This proceeds through collisionless interactions, and thus the fine-grained phase-space distribution of galaxies is conserved. However, if the relaxation is sufficiently violent (for example, if the energy of stars is substantially changed by a fluctuating potential), then initially adjoining units of phase-space become widely separated in the final state through phase mixing. Averaging the fine-grained phase-space density over any observable volume gives a coarse-grained distribution, which can be an equilibrium distribution. 

Donald showed by an ingenious argument in statistical physics that the endpoint of violent relaxation is independent of the details of the initial state. In doing this, he discovered a new type of statistical mechanics. Particles that are indistinguishable follow Fermi-Dirac or Bose-Einstein statistics according to whether they obey an exclusion principle or not. Particles that are distinguishable without an exclusion principle obey Maxwell-Boltzmann statistics. The remaining case of distinguishable particles obeying an exclusion principle is now known as~\lq\lq Lynden-Bell statistics" in standard texts~\citep{Ha77}. It provides the equilibrium distribution function of a collisionless system in the idealized limit when violent relaxation proceeds to completion.

Nowadays, the Lynden-Bell distribution function is not believed to describe the equilibrium properties of elliptical galaxies. Since the 1990s, increasingly powerful numerical simulations have mimicked the process of violent relaxation as part of the recipe of galaxy formation. Violent relaxation is more ineffective than envisaged in Donald's statistical argument, and some memory of environment and initial conditions is retained. Donald recognized this shortcoming in later work on $H$-functions (\ref{THL}) with Scott Tremaine FRS 1994 and H\'enon (THL). Here, the aim was to constrain the direction of evolution of a collisionless system by the principle of increase in entropy, a more modest goal than in (\ref{LB1967}). THL showed any equilibrium stellar system can be described as the stationary point of some $H$-function -- analogous to Boltzmann's $H$-function in the kinetic theory of gases -- provided its phase space distribution function is a decreasing function of energy alone. All such $H$-functions increase during violent relaxation, and the distribution function becomes more and more mixed. THL used this to demonstrate a violently relaxed galaxy resembles observed elliptical galaxies only if the initial state is cold and clumpy.
 
\subsection{The Gravothermal Catastrophe and Negative Heat Capacities}

Donald received through the mail an astonishing 1962 paper by Vadim Antonov from then Leningrad in the Soviet Union. It was finally translated and made widely available in English over twenty years later~\citep{An85}. Donald was unable to read Russian, but he could follow the mathematics. Antonov took an assembly of self-gravitating particles in a spherical container and maximized their entropy at fixed total energy. He found that the entropy was only a (local) maximum if the ratio of the density at the centre to that at the container is less than 708. Above this value, no maximum exists. Donald carried out his own investigations of the thermodynamics of a self-gravitating gas in a spherical container in collaboration with Roger Wood~(\ref{LB1968}). At a density contrast of 389, Donald found that the heat capacity of the gas at constant volume becomes infinite and, for slightly greater contrasts, it attains large negative values. Henceforth, the heat capacity increases as the contrast increases, finally reaching zero at Antonov's critical value~\citep{BT}. Donald discovered that there is a range of density contrasts from 389 to 708 at which maximum entropy systems exhibit negative heat capacity. Although it had long been known from the virial theorem that negative heat capacities occur for isolated self-gravitating systems, the result was believed to have limited implications. Isolated systems cannot be in thermal equilibrium under gravity because particles can escape through evaporation or encounters. Antonov's and Lynden-Bell \& Wood's gas spheres in containers were the first true equilibrium systems displaying negative heat capacity. 

Donald quickly realised that this was the basis for an runaway instability, which he coined \lq\lq the gravothermal catastrophe". At high temperature, a negative heat capacity system in contact with a cold sink will give up heat until it is entirely exhausted, getting hotter, rather than colder, as it does so. So, if a high temperature and a low temperature gravitational system are in contact, energy is transferred out of the hot portion, which contracts, and into the cold portion, which expands. This naturally builds self-gravitating systems with tight cores and extended haloes or coronae. Donald originally envisaged applications to the contracting phase of a red giant star. We now believe that the catastrophe occurs when the core of a globular cluster shrinks and heats up, causing it to transfer energy to stars in the cluster's halo, and leading to core collapse. Donald, in later work with Peter Eggleton, followed up the implications of the gravothermal catastrophe for globular clusters~(\ref{LB1980}). They showed not just the evolution of the core, but also the surroundings, are self-similar.  The debris of material expelled from the core follows a power law in the density with exponent in the range -2 to -2.5.

Negative specific heats fascinated Donald and he returned to the subject repeatedly. There is a standard thermodynamic proof that heat capacities are always positive, repeated by \citet{Sc68} in {\it Statistical Thermodynamics}. So, Donald's ideas were initially received with scepticism by theoretical physicists. It was later demonstrated by \citet{Th70} that the standard proof only holds for systems that can be in equilibrium in canonical ensembles. Microcanonical systems can have negative heat capacities. Together with Ruth Lynden-Bell, Donald showed that phase transitions may be viewed as being due to microscopic elements with negative heat capacity~(\ref{LB1977},\ref{LB1999}). He identified such units in the theories of ionization, chemical dissociation and the Van der Waals gas. Donald therefore believed that negative heat capacities are of general applicability in physics, and not confined to self-gravitating bodies, like star clusters and black holes.
 
\subsection{Galactic Stability, Orbital Cooperation and Spiral Structure}

Donald started by studying cooperative phenomena in homogeneous stellar systems~(\ref{LBCoop}). Although not realistic, such models can still do useful work (as Chandrasekhar's (1942) calculation of the dynamical friction formula in {\it Principles of Stellar Dynamics} showed). Donald proved that two-stream instabilities do not exist in a homogeneous stellar system consisting of two Maxwellian streams. Using Nyquist diagrams, he found that instability occurs only when at least one of the streams has a flat-topped velocity distribution. With Nigel Sanitt, Donald showed that a stellar system is stable whenever the corresponding barotropic gaseous system is secularly stable~(\ref{LBS}). Their Schr\"odinger operator method (essentially an energy principle) showed that a large class of spherical stellar systems are stable to all non-spherical modes of vibration, but stability to spherical modes was frustratingly harder to assess. Donald concluded, sadly but correctly, that~\lq\lq more general and more powerful methods are needed in the field of galactic stability". In fact, progress in identifying the instabilities of spherical or elliptical galaxies has not come from energy principles, but had to wait until the advent of powerful numerical simulations in the 1990s. 

Donald also developed a secular stability criteria for stars and galaxies in work with Jeremiah Ostriker, then an NSF Fellow in Cambridge. In their paper (henceforth LBO), they derived a variational principle to determine the secular stability of any differentially-rotating, self-gravitating fluid flow. With some modification~\citep{BFSS}, this has proved useful in many areas of astronomy -- for example, helioseismology, the stability of rotating stars and the effects of tides in extrasolar planets. LBO applied their result to the normal modes of oscillation in galactic discs and formulated the \lq\lq Anti-spiral Theorem". This states that, if in linear theory there exists a global spiral mode of trailing planform that rotates without growing or decaying, then a similar mirror-image leading mode must also exist as well. The Anti-spiral Theorem was criticism of the then prevalent idea of long-lived quasi-static spiral structure promulgated by C.C. Lin and co-workers. This was the very antithesis of GLB's view of spirality as sheared gravitational instabilities. Donald remarked later:~\lq\lq I always held the view that angular momentum transfer is the driving force behind spiral structure. In part the Anti-spiral Theorem was there because it seemed to point out that what C.C. Lin said was much less than the whole story."

Donald returned to the fretfulness of spiral structure theory when Agris Kalnajs arrived as a Research Fellow at the RGO. Together, Lynden-Bell \& Kalnajs (LBK) laid one of the cornerstones of stellar dynamics, studying the role of resonances in the growth and decay of spiral waves~(\ref{LBK1972}). LBK showed that there is absorption of angular momentum by stars that resonate with the wave at the outer Lindblad resonance. Emission of angular momentum occurs at the inner Lindblad resonance. LBK derived a formulae for angular momentum transport and showed that the role of spiral structure is to carry angular momentum from the inner to the outer parts. LBK likened this advective transport to a fleet of trucks carrying coal. The trucks travel outward full of coal and return empty, so there is an outward flow of coal but no net flow of trucks. Similarly, stars can carry angular momentum outward, deposit it near the apocenter of their orbits, and return to acquire more angular momentum near pericenter, leading to an outward flow of angular momentum with no outward flow of mass. LBK also noted that an orbiting star subject to weak tangential forces can exhibit unusual behaviour. Orbits close to corotation have very small mean motions in the corotating frame. Those on the inside drift slowly forwards, those on the outside drift slowly backwards. On feeling the forward tug of a spiral arm, a forward moving star gets onto an epicycle with a slightly greater angular momentum and its mean motion, far from speeding up, will slow down. Thus, uncooperative stars act \lq\lq like donkeys slowing down when pulled forwards and speeding up when held back." Donald extended these ideas into a lucid treatment of resonant orbits~(\ref{LB73}), borrowing techniques that Max Born FRS 1939 originally developed in {\it The Mechanics of the Atom} during the days of the old quantum theory.

LBK also suggested that galactic bars are standing waves which orient and trap the major axes of orbits with two lobes so that they lie along the bar. (This is physically identical to the radial orbit instability in elliptical galaxies). In this picture~(\ref{LB79bar}), bars can form only in regions in which the rotation curve rises slowly to a maximum. Once the rotation curve turns over, the major axis of an elliptic orbit behaves like a \lq donkey'. When subject to a torque pulling forward, its precession slows down, so it anti-aligns with the bar's potential well and weakens it. This elegant idea may hold true for some slow bars, but it is not the whole story, as numerical simulations and observational data both favour fast bar pattern speeds. As in the Jeans instability, gravity will have its way in organising orbits only provided the system is not too hot. So the mechanism of aligning and anti-aligning orbits has to be taken hand-in-hand with the population of orbits. Just as a hot gas can resist Jeans collapse, so a wide dispersion of orbital precessions can inhibit the formation of structures like bars through orbital cooperation.

\subsection{Black holes as quasars}

The most famous piece of research that Donald did at Herstmonceux Castle is his startling idea that quasars are powered by supermassive black holes and that most large galaxies, including our own, could host a dead quasar in their nucleus~(\ref{LB1969}). Years later, this was to win him the 2008 Kavli Prize in Astrophysics, jointly with Maarten Schmidt, \lq\lq for their seminal contributions to understanding the nature of quasars".

Quasars were identified as ultraluminous active galactic nuclei following Cyril Hazard's accurate radio position of the quasar 3C 273 in 1963, followed by Maarten Schmidt's determination of its redshift of 0.16. The physical mechanism that powered the quasars remained obscure. The nuclear region of 3C 273 is then less than 1 kpc in diameter, whilst its associated radio and optical jet is about 50 kpc away, implying a timescale greater than $10^5$ years. The optical luminosity of 3C 273 is enormous, at least 10 times brighter than the brightest giant elliptical galaxy. The idea that black holes may provide the energy source for quasars had been mooted in the literature -- though the term \lq\lq black hole" only became commonplace after 1967. \citet{HF63} suggested that "only through the contraction of a mass of $\approx 10^8 M_\odot$ to the relativistic limit can the energies of the strongest sources be obtained". Soon after, \citet{Salp64} and \citet{Ze64} proposed that accretion onto a supermassive black hole may be the source of the luminosity. If material gradually spirals on to the innermost stable orbit of a nonrotating black hole at $r = 6GM / c^2$, the energy released per unit mass is $\approx 0.06 c^2$, enough to provide the energy of a luminous quasar from a plausible mass.

Donald was aware of the frenzy amongst observers about quasars from his time as Harkness Fellow at CalTech. A freak of numerology now spurred him on. \lq\lq It was now four years since I had left Cambridge for the RGO and from our home in Barcombe I drove to Herstmonceux with Bernard Pagel FRS 1992. Part of our route lay along the A273 road, and this daily reminded me that there was still no accepted theory of the quasar 3C 273"~(\ref{LB2010}). Donald knew that quasars have active periods and only rarely are they very bright. He made a rough estimate of how common dead quasars would be and concluded that the nearest one to an average point in the Universe would be about 3 Mpc away. He then argued that dead quasars in the form of~\lq\lq collapsed bodies" (or supermassive black holes) should be ubiquitous in galactic nuclei, given the lifetime energy output of quasars and their abundance at early times. He suggested that dead quasars may be detectable through measurement of the mass-to-light ratios of nearby galactic nuclei. Donald explored the thermal radiation and particle emission produced by a disk of gas orbiting the hole, with energy dissipation related to magnetic and turbulent processes. He found that~\lq\lq with different values of the [black hole mass and accretion rate] these disks are capable of providing an explanation for a large fraction of the incredible phenomena of high energy astrophysics, including galactic nuclei, Seyfert galaxies, quasars and cosmic rays." This very influential paper made plausible to most astronomers that the black hole model -- which had hitherto received scant attention -- was the correct explanation of quasars and active galactic nuclei. Together with Martin Rees, he explored the idea that the nearest supermassive black hole to us may even be at the centre of the Milky Way Galaxy~(\ref{LBR}).

Donald's predictions were confirmed within his lifetime. High resolution spectroscopy, especially with the {\it Hubble Space Telescope}, of galactic nuclei produced an abundance of evidence for compact dark mass concentrations of up to $10^8 M_\odot$ pc$^{-3}$~\citep{Ko95}. Even stronger evidence exists for NGC 4258, for which the rotational kinematics of water masers show the existence of a central supermassive black hole with mass of $3.9 \times 10^7 M_\odot$ ~\citep{Mi95}. The black hole at the centre of the Milky Way was confirmed by spectroscopic and proper motion studies of stars, exploiting adaptive optics~\citep{Sc02, Gh08}. The compact radio source Sagittarius A$^\star$ is the Milky Way's supermassive black hole. It has a mass $\approx 4 \times 10^6 M_\odot$ and a Schwarzschild radius of $0.08$ astronomical units, as judged from the accelerations of stars orbiting around it. Final confirmation of Donald's prediction was provided by the {\it Event Horizon Telescope}, which synthesized radio images of M87 to give a dramatic picture of the shadow of its central supermassive black hole -- perhaps the single most famous image of 2019, but one that Donald did not live to see.

\section{Return to Cambridge (1972)}

Sir Richard Woolley retired as Director of the RGO in 1971 and this precipitated Donald to look for a professorship elsewhere. Donald was particularly interested in developing his ideas in mathematical relativity, and so wanted stimulus and encouragement from other theoreticians, whereas the RGO was almost wholly observational. As chance would have it, two astronomical professorships were vacant in Cambridge, the Jacksonian Chair in Natural Philosophy at the Cavendish Laboratory and the Chair of Astrophysics at the Observatories, formerly occupied by the observational astronomer Roderick Redman. Donald consulted Ryle on which position to apply for. Ryle recommended Donald apply for the Chair of Astrophysics.
 
What then happened is entangled in the confusing story of the enmity between Ryle and Hoyle, then the Plumian Professor of Astronomy and Natural Philosophy at the Institute of Theoretical Astronomy, Cambridge. Donald's recollection of events is given in (\ref{LB2010}), whereas Hoyle's lively memories are recounted in his rollicking autobiography, {\it Home is Where the Wind Blows}~\citep{Ho94}. Although Donald behaved with good grace throughout, events sadly so conspired as to cause Hoyle to resign his Professorship and leave Cambridge forever.

Unbeknownst to Donald, Hoyle had hoped to encourage a leading observational astronomer to occupy the Chair of Astrophysics. His preferred candidates were Sargent or Geoffrey Burbidge FRS 1968. The other electors dissented, and after much wrangling, Donald was elected against Hoyle's wishes.

This all occurred whilst astronomy in Cambridge was undergoing reorganization. Hoyle had founded the Institute of Theoretical Astronomy (IoTA) in 1968, thanks to grants from Lords Wolfson and Nuffield. These provided funds for a building, a powerful computer, and 12 posts for five years. Hoyle attracted many young and gifted astronomers to IoTA, but the long-term funding was always precarious. Anticipating the expiry of the starting grants, the University produced a report on the future of Cambridge astronomy, which advocated that IoTA be merged with the old Observatories. This was very much Hoyle's vision, as he hoped Cambridge could complement its strong theoretical research with observational programs exploiting new large-scale facilities, like the Anglo-Australian Telescope. This merger created a new University department, the present-day Institute of Astronomy (IoA).

Hoyle assumed that he would be the Director of the newly created IoA, to which Donald was happy to assent having no great desire to take on administrative duties~(\ref{LB2010}). However, in a situation of muddled complexity, Hoyle came to believe that Donald was being chosen as the Director behind his back by the University. Hoyle perceived this as the culmination of many slights, following years of prolonged feuding with some senior members of the British astronomy community~\citep{Ho94}. He resigned from the Plumian chair in September 1971 and left Cambridge academic life forever for the remote Cockley Moor in the fells above Ullswater.

So, when Donald arrived in Cambridge in 1972, he faced a huge task as director of a department which had been created under disputatious conditions and which had 12 staff members whose jobs would expire at the end of the year! Donald's pragmatic reaction was to strike a deal with George Batchelor FRS 1957, fluid dynamicist and Head of the Department of Applied Mathematics and Theoretical Physics (DAMTP). Mathematical relativity and theoretical cosmology were henceforth transferred to DAMTP, who provided permanent lectureships. This was the nucleus of DAMTP's general relativity and gravitation group, which was to became world-famous under Stephen Hawking's intellectual leadership in the 1980s and 1990s. Two further lectureships, one shared with DAMTP, were created and offered to Douglas Gough FRS 1997 and Peter Eggleton. The new department was beginning to stabilize, when, in a bold coup, Donald enticed Martin Rees FRS 1979 to return from the University of Sussex as a youthful Plumian Professor. Donald and Martin were to run the IoA amicably over the next twenty years, alternating the Directorship for five year periods from 1972 to the mid 1990s. The number of staff, fellows and students increased from 70 in 1972 to 138 at the end of Donald's reign as Director in 1994. Donald and Martin together oversaw the growth of the IoA into one of the premier research institutes in astronomy in the world.

Despite a difficult induction as Director, Donald was generously forgiving. His opinion of Hoyle's many scientific achievements remained extremely high~(\ref{LBObitHoyle}). Donald always regarded Hoyle as possessing a very remarkable combination of scientific imagination, mathematical ability and physical insight (as well as Yorkshire bluntness). He strongly regretted the omission of Hoyle from the award of the Nobel Prize in Physics in 1983, which went to Hoyle's long-time collaborator William Fowler for studies of the formation of the chemical elements in the universe. The author recalls Donald remarking forcefully,~\lq\lq The work would never have happened without Fred".

\begin{figure}
\includegraphics[width=\textwidth]{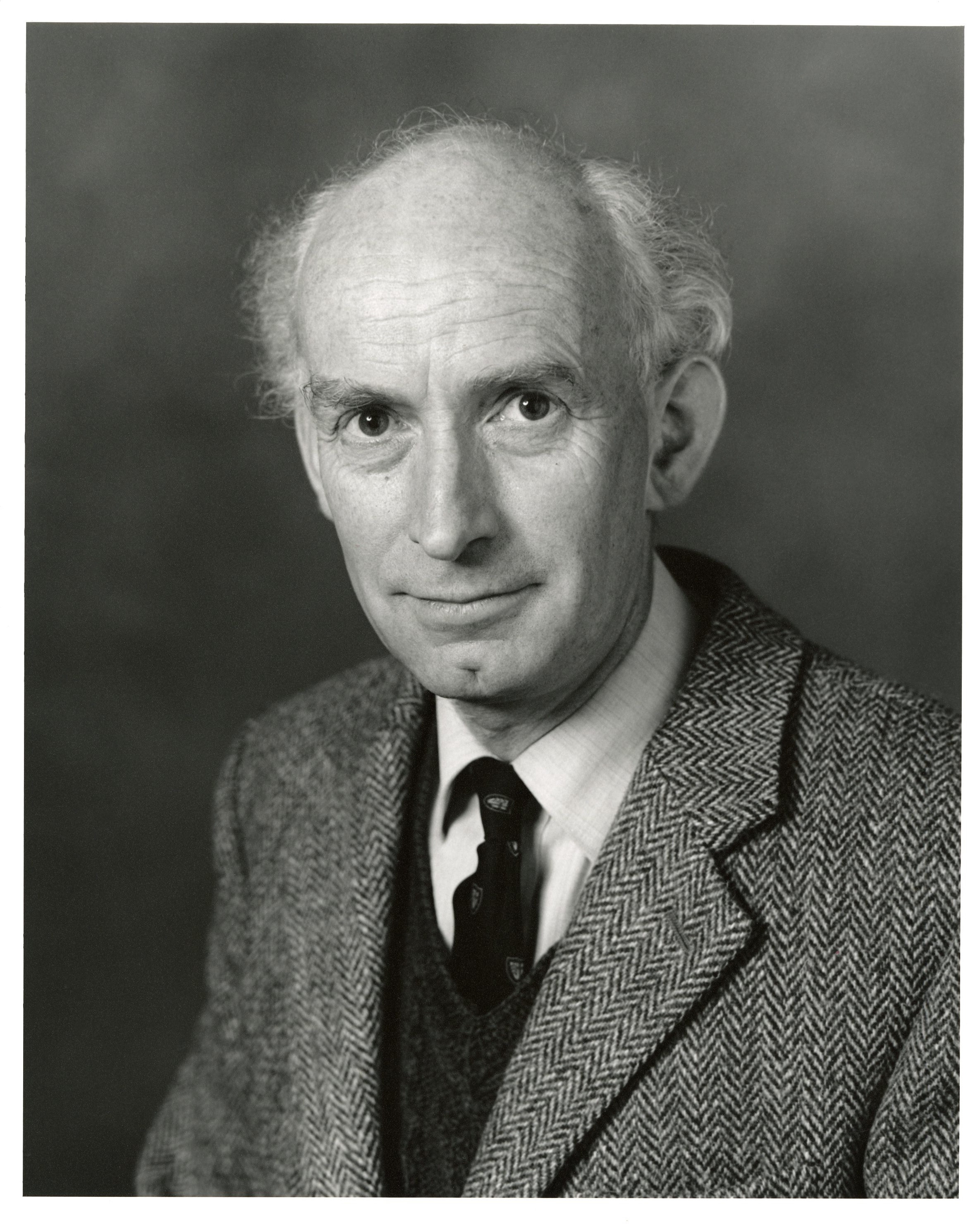}
\caption{Donald Lynden-Bell during the 1970s while Professor of Astrophysics at Cambridge University (Courtesy of Ruth Lynden-Bell).}
\label{fig:dlbconf}
\end{figure}

\section{Professor of Astrophysics, University of Cambridge (1972-1996)}

\subsection{Angular Momentum Transport in Accretion Discs}

Nowadays, there is huge interest in accretion disks, driven by their ubiquity in astrophysics from supermassive black holes in active galactic nuclei to star and planet formation. This has ensured that Donald's next paper (\ref{LBP74}) is now his most highly cited of all.

Given a source of dissipation, an accretion disc causes angular momentum to be transported outward, leading to an expansion of the outer parts, accompanied by the accumulation of more and more mass towards the centre. This observation is at least as old as 1927, appearing in Sir Harold Jeffreys' discussion of the solar nebula in the first edition of his textbook {\it The Earth}. In his {\it Nature} paper on quasars, Donald assumed that the main mechanism was magnetic viscosity, or the transfer of angular momentum by magnetic stresses. But, Donald had forgotten his own earlier work! In his 1960 Ph.~D. thesis, he had already derived -- but not published -- a set of solutions for thin accretion discs with kinematic viscosity. A graduate student at the IoA, Jim Pringle then re-discovered them in his 1974 Ph.~D. thesis! And, although they did not know it then, both had been partly anticipated by Rainer \citet{Lu52}, a student of Carl von Weizs{\"a}cker~(see \citet{Pr81} for the historical details). 

Together, Donald and Pringle (hereafter LBP) joined forces to write an important paper on accretion discs~(\ref{LBP74}). The equation for the evolution of the surface density of an disc admits exact analytic solutions when the kinematic viscosity depends only on a power-law of the radius, and not on time or surface density. The solutions fall into two classes, one for which the torque at the centre is zero and all the matters flows to the centre (accretion discs), the other for which the torque at the centre is finite, and the matter is expelled to infinity (decretion discs). LBP also provided the Green function -- that is, the evolution of a viscous ring of material -- from which these more general solutions of the linear parabolic equation are constructed. Turbulent flows and the resulting eddies certainly offer a transport mechanism for angular momentum and mass, which is independent of the existence of magnetic fields. LBP provided a generic model for the evolution of thin accretion discs, though detailed specification of the angular momentum transport and any mass loss processes at work is needed for predictive power in any particular application.
 
The main application in LBP is to what were then called \lq\lq the nebular variables", nowadays T Tauri or pre-main sequence stars. At the time, it was known that T Tauri stars have both an ultraviolet excess and an infrared excess. LBP's explanation is that the excesses are caused by accretion discs around these young stars. The flux from the disc has a flatter spectrum than Rayleigh-Jeans and so dominates the spectrum at longer wavelengths and provides the infrared excess. The blue end is dominated by emission from the boundary layer where the disc and star meet and high temperatures prevail. With some modifications, this has remained the standard picture till today. The infrared excess is one of the primary methods used to identify young stellar objects and proto-planetary discs in astronomical surveys.

\subsection{The Large-Scale Streaming of the Galaxies}

In the 1980s, Donald was one of seven astronomers -- the Seven Samurai
-- who conducted an ambitious program to measure the distances and
recession velocities for 400 elliptical galaxies. The aim was to
understand the large scale peculiar motions of nearby galaxies
(including our own) with respect to the Hubble flow. The other Samurai
were Professors Sandy Faber, Roger Davies, David Burstein, Alan
Dressler, Roberto Terlevich and Gary Wegner. The Samurai devised a new
distance indicator for elliptical galaxies~(\ref{Dr87}), the so-called
$D_n-\sigma$ relation where $\sigma$ is the central velocity
dispersion and $D_n$ is the diameter of the central region within
which the average surface brightness exceeds 20.75 mag
arcsec$^{-2}$. This distance estimator was a substantial advance on
the earlier Faber-Jackson relation, but nowadays it has itself been
superseded by the Fundamental Plane~\citep{BM}. Unexpectedly, the
Samurai discovered a coherent, large amplitude flow in the direction
of Hydra-Centaurus, which they named the \lq\lq Great Attractor''~(\ref{LB88}). It is situated at a distance of roughly 80 Mpc away from the Milky Way in the direction of the constellation Norma. The Great Attractor is pulling in millions of galaxies in a region of the Universe that includes the Milky Way, the Local Group and the larger Virgo Supercluster, and the still larger Hydra-Centaurus Supercluster, at velocities of up to a thousand kilometers per second~(\ref{Dr87},\ref{LB88}). Based on the kinematics, the unseen mass inhabiting the voids between the clusters of galaxies was estimated to be $\approx 10$ times more than the visible matter. The total mass of the Great Attractor was reckoned as $\approx 5 \times 10^{16} M_\odot$. 

At the time, this was one of the most ambitious galaxy redshift surves ever attempted. It led to the discovery of bulk motions comparable in magnitude to the motion of the Local Group through the cosmic microwave background (CMB). Together with Ofer Lahav, Donald showed that the optical dipole lies within 7$^\circ$ of the direction of the Local Group's motion through the CMB. The directions of the optical, infrared and CMB dipoles are all consistent with each other~(\ref{LLR}). Donald, Lahav and others used this for cosmological parameter estimation~(\ref{LLB89}), though this work predates the discovery of the non-zero cosmological constant ${\rm \Lambda}$. Calculations of the mass of the Great Attractor have subsequently been revised downward by about an order of magnitude following infrared and X-ray studies~\citep{Ko07}. Galaxies located on the other side of the Great Attractor are no longer thought to be pulled in its direction, as further more massive agglomerations of galaxies lie behind it. These comprise the Shapley Supercluster, about four times more distant than the Great Attractor~\citep{Ko06}. 

Galaxy surveys proved to be a productive research area over the next decades, but the field took a different turn. With the development of plate scanning machines, computers replaced humans in identifying galaxies, as in the APM Galaxy Survey. Using this as an input source catalogue, the Two-Degree Field Galaxy Survey obtained redshifts for $\approx$ 220\,000 galaxies. These were used to measure the two-dimensional two-point correlation function, from which the projected and redshift space correlation functions are derived. Donald continued to be interested in the area, especially in techniques for recovering the cosmological density, velocity and potential fields from all-sky redshift catalogues. He developed a method based on expansion of the fields in orthogonal basis functions~(\ref{FLB}). Peculiar velocities introduce a coupling of the radial harmonics describing the density field in redshift space, but leave the angular modes unaffected. In linear theory, the radial coupling is described by an analytic distortion matrix which can be inverted to give the real space values. Statistical or `shot' noise is mitigated by regularizing the matrix inversion with a Wiener filter. 

The notion that there may be galaxies hidden behind the Zone of Avoidance, where dust and gas in the Milky Way obscure about a quarter of the Sky, was also fruitful. It inspired Donald to participate in HI surveys of this region. This led to the discovery of a large barred spiral galaxy, Dwingeloo 1, about 3 Mpc away~(\ref{LBDwin}). It is a member of a group containing IC 342 and the Maffei galaxies, just beyond the Local Group in which the Milky Way resides.

\subsection{Tidal Streams, Dwarf Galaxies and Near-Field Cosmology}

By the late 1970s, it was clear that the monolithic collapse model for the formation of the Galaxy in ELS (\ref{ELS}) was incorrect. The idea that mergers were major players in the ongoing drama of the formation of the Milky Way was directly confirmed when first the Magellanic stream~\citep{Ma74}, and then the Sagittarius stream~\citep{Ib94}, were discovered. 
It is hard to improve on Donald's own description of the curiosity that propelled him forward: \lq\lq Some bad weather during an otherwise successful photometric observing run at SAAO, Sutherland, gave me time to ponder the Magellanic stream and to wonder whether, like the Magellanic Clouds, other near neighbours of our Galaxy were associated with streams of neutral hydrogen. So, taking C. W. Allen's {\it Astrophysical Quantities}, I plotted the small satellites of the Milky Way on to R. D. Davies' recent map of high velocity hydrogen. Taking our satellites in order of distance I found that the Large and Small Magellanic Clouds, Draco, Ursa Minor and probably Sculptor all lay in or close to the directions of high velocity hydrogen streams"~(\ref{LB1976}). The association of Draco and Ursa Minor with the Magellanic Stream was strengthened by the discovery that their elongations lie along the line of the Magellanic Stream~(\ref{LB1982}), also noted by~\citet{Hu77}. Donald believed the fact that that Sculptor, although lying close to the Magellanic Stream, nevertheless points at Fornax, and that this line passes on through the distant satellites Leo I and II, suggests that there may be another tidal stream, distinct from the Magellanic Stream~(\ref{LB1982},\ref{LB1983}). In this picture, the dwarf spheroidal satellites of the Milky Way lie on two streams and are elongated along them~(\ref{LB1983}).

Donald quickly realised the importance of tidal streams for measurements of the enclosed mass. He embarked on a suite of computer simulations of the Magellanic system with Doug Lin, then a hungry research fellow. The details of the simulations have not survived the test of time, as Donald assumed that the Magellanic Stream is produced by tidal interaction of the neutral hydrogen envelope of the Large and Small Magellanic Cloud with the Galaxy. Nowadays, with the benefit of accurate proper motions, we know that interactions between the Large and Small Magellanic Clouds created the Stream. This though does not lessen the significance of the first mass measurement with tidal streams. The conclusions (\ref{LL82}) that the Galaxy must have a massive dark halo out to at least 70 kpc and that the circular velocity of the Galaxy is $244\pm 20$ kms$^{-1}$ are completely correct.

The discovery of the Sagittarius stream in 1994 fired up Donald's interest anew. Together with Ruth Lynden-Bell, he argued that objects belonging to the same tidal stream lie on the same great circle in the Galactocentric sky~(\ref{LBLB1995}). This is exactly true if the stream follows an orbit in a spherical potential. The Galactocentric direction to any object then defines the pole of a great circle. So, the great circle joining any two objects has as its pole the intersection of the two great circles whose poles are at the objects. Thus, if there are many objects belonging to a stream, there will be a multiple intersection point of all the great circles whose poles point at the objects. Donald searched for such intersections and was able to recover the earlier streams associated with the Magellanic Clouds and with Fornax, and to identify new ones associated with Sagittarius and its former globular clusters. The {\it Sloan Digital Sky Survey} provided multiband photometry on Galactic stars covering a quarter of the sky around the Galactic North Pole. This dataset proved to be a treasure trove in the hunt for streams. When Vasily Belokurov and Wyn Evans discovered a delicate track of stars crossing~\lq\lq the Field of Streams", they christened it~\lq\lq the Orphan Stream" because of its lack of obvious progenitor. Donald took an immediate interest in this parentless stream~(\ref{Be07}). He soon pointed out an astonishing fact. The track of the Orphan Stream on the sky exactly matches a distended high velocity stream of neutral hydrogen, Complex A. Subsequently, Donald showed that a solution exists in which the two streams share the same orbit, though they lie on different wraps~(\ref{JL}).

Donald was the first to see the great future of tidal streams as Galactic potentiometers. He assumed streams are orbits, enabling him to use his favourite workhorse of analytic methods. This has been superseded by numerical codes that generate streams using Lagrange point stripping~\citep{Gi14}, revealing that streams are not exact orbits. In fact, deviations from orbits provide some of the most interesting results -- such as the recent measurement of the mass of the Large Magellanic Cloud through its subtle perturbations on the Orphan Stream that twist the latter's orbital plane~\citep{Er19}. Donald's work on the planes of dwarf satellites around the Milky Way has also evolved into a field of heated controversy. We now know that the distribution of satellites around the Milky Way is highly anisotropic and that the Magellanic Clouds are surrounded by their own retinue of dwarf galaxies~\citep{Kop}. Whether this anisotropy reveals fundamental flaws with our ideas of galaxy formation or whether such anisotropies follow from group infall of satellites along preferential directions in the cosmic web remains a contested problem today.

\begin{figure}
\includegraphics[width=\textwidth]{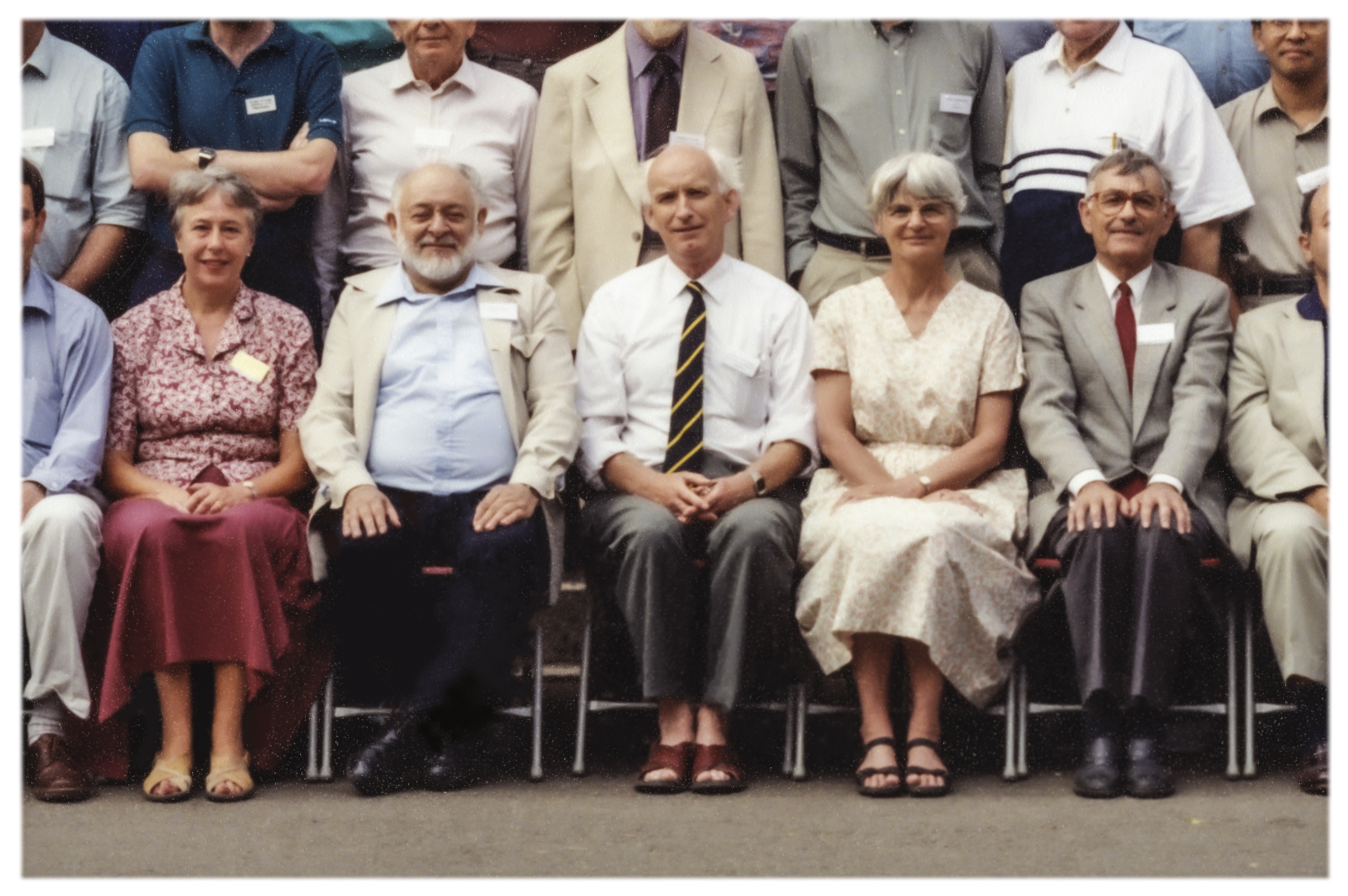}
\caption{Donald Lynden-Bell at the conference on \lq Gravitational Dynamics' in honour of his sixtieth birthday held in Cambridge in 1995. Shown left to right are Alice Duncan (Donald's secretary), Prof
Leon Mestel (Donald's PhD supervisor), Donald himself, Prof Ruth Lynden-Bell and Prof Joseph Katz
(Donald's long term collaborator on relativity). (Courtesy of Lafayette Photography.) }
\label{fig:dlbconf}
\end{figure}

\subsection{Mach's Principle and General Relativity}

Joe Vinen, his tutor at Clare College, encouraged Donald's \lq\lq first groping steps" to an understanding of inertia in an undergraduate essay~(\ref{LB1967}). Donald liked the formulation of Mach's principle in one of his favourite books, {\it Cosmology} by \citet{Bo52}, who wrote \lq\lq Local inertial frames are determined through the distributions of energy and momentum in the Universe by some weighted averages of the apparent motions". This led him to study the associated effect of~\lq\lq dragging of inertial frames", or~\lq\lq gravomagnetism". The main consequence of the gravomagnetic field (or velocity-dependent acceleration) is that a moving body near a massive rotating object experiences an acceleration not predicted by Newtonian gravity. There are further subtle predictions, such as induced rotation of a falling object and precession of a spinning object. At the RGO, Donald already introduced a reformulation of the Einstein field equations as integral equations involving retarded Green's functions in Friedmann-Robertson-Walker (FRW) universes~(\ref{LB1967}). This approach stalled, because it did not attribute any inertial influence to gravitational waves. 

Once in Cambridge, general relativity became one of his favourite research pastimes. He struck up a productive and long-standing collaboration Joseph Katz and Jiri Bi\v{c}\'{a}k. They showed that, within the context of linearized perturbation theory of FRW universes, Mach's principle follows from the constraint equations of general relativity, provided that the universe is closed~(\ref{LB1995KB}). They studied frame dragging effects in a number of situations, such as inside a collapsing and slowly rotating spherical matter shell, or inside rotating cylinders and investigated how rotations beyond the cosmological horizon affect the local inertial frame. They introduced~\lq\lq Machian gauges" to study general linear perturbations of FRW universes~(\ref{BKL}). These admit much less freedom than the synchronous gauges commonly used in cosmology, but they allow local inertial frames to be determined instantaneously via the perturbed Einstein field equations from the distributions of energy and momentum in the universe. This is directly inspired by Donald's understanding of Mach's principle. The study of Machian effects inspired the formulation of conservation laws with respect to curved backgrounds and the introduction of the Katz-Bi\v{c}\'{a}k-Lynden-Bell (or KBL) superpotentials~(\ref{KBL}). These are unambiguous in spacetimes with or without a cosmological constant, and have found numerous applications -- for example, in the studies of the back reactions in slow-roll inflation or in finding energy-mass and other total charges for black holes in asymptotically non-flat backgrounds. The construction of the KBL superpotentials is Donald's most cited paper in general relativity.

Donald realised that an infinite disc with a flat rotation curve can additionally only involve the constants of general relativity -- Newton's gravitational constant and the speed of light. No length scale can be made out of these constants, so the geometry is self-similar. Together with Serge Pineault, he constructed elegant solutions to the field equations for self-similar counter-rotating discs, as well as numerically computed rotating solutions~(\ref{LB1978a},\ref{LB1978b}). At the end of 1980s, Donald studied spherical self-similar solutions for cold collapse with Jos\'e Lemos~(\ref{LL88},\ref{LL89}). 
The main achievement was the detailed illustration of self-similar Newtonian solutions for non-crossing, collapsing spherical shells of matter. In the first paper, each shell was assumed to fall from rest at infinity. Next, self-similar solutions for collapsing and expanding non-crossing shells were considered. 
They analyzed the general relativistic analogues of the Newtonian solutions, which are the Lema\^{i}tre-Tolman-Bondi metrics for dust. 
Among known finite-mass non-spherical solutions of Einstein's equations, few have physical sources. From 1992, Donald worked on the idea of thin relativistic discs as possible sources of static spacetimes. Together with Katz and Bi\v{c}\'{a}k, he showed that most vacuum Weyl solutions can arise as the metrics of counter-rotating relativistic discs~(\ref{BLKdisc}). Inspired by new developments in Newtonian potential theory~\citep{EZ}, he also constructed an infinite number of new static, relativistic disc solutions, including disc sources for the Kerr metric~(\ref{BLP}). Many of Donald's exact solutions to the Einstein field equations have found their way into the standard reference book~\citep{Mac}.

Although Donald's work in relativity was not as influential as his work in astrophysics, it gave him enormous pleasure. He did not favour the geometric approaches to general relativity common in many modern textbooks, but much preferred the physical treatment in the celebrated book by Landau \& Lifshitz on {\it The Classical Theory of Fields}, repeatedly extolling it as~\lq\lq the best book on relativity ever written".

\begin{figure}
\includegraphics[width=\textwidth]{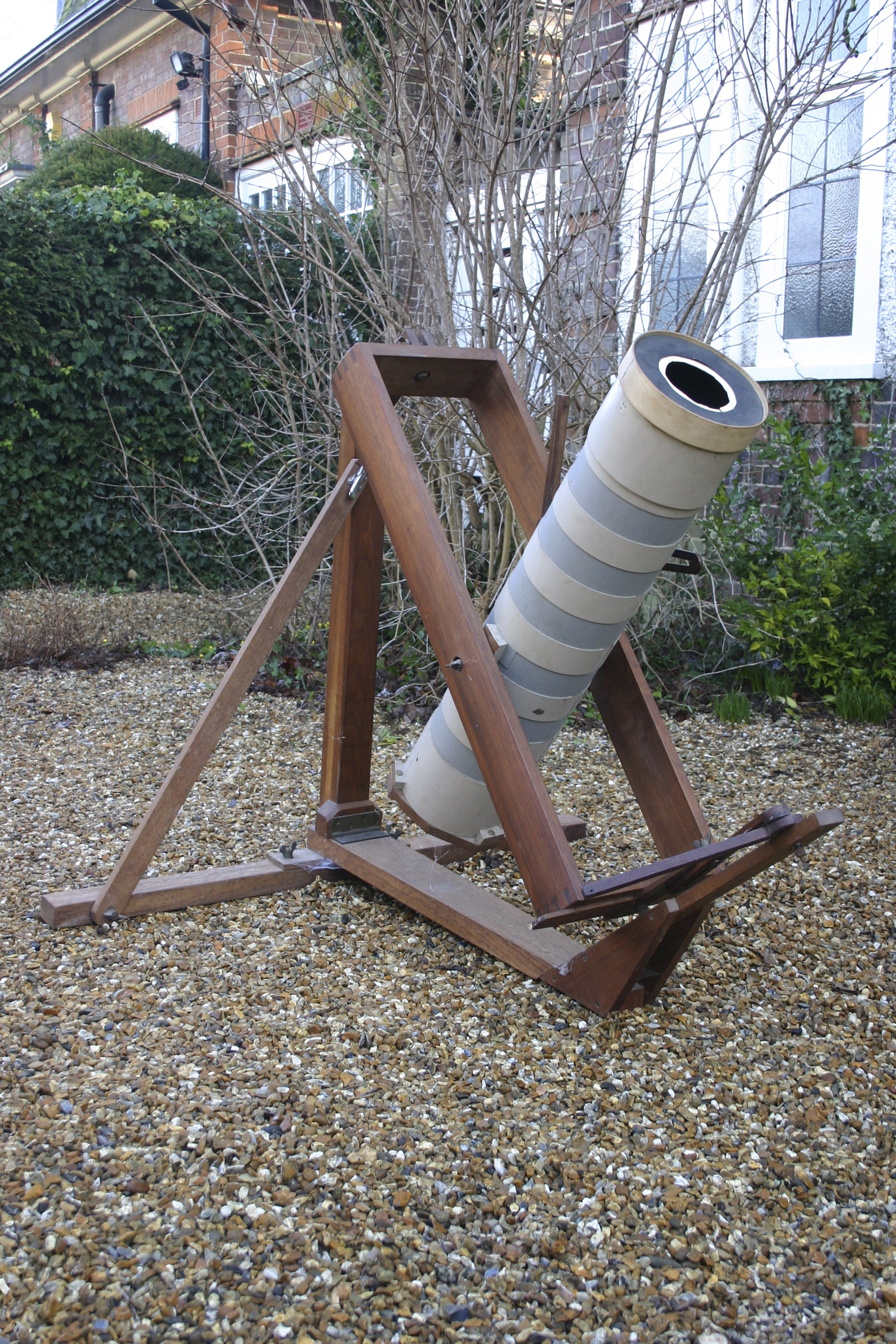}
\caption{Donald's telescope. Its mirror was constructed by Donald as an undergraduate. (Courtesy of
Amanda Smith.)}
\label{fig:dlbtel}
\end{figure}

\section{Belfast, Cambridge (1996-2018)}

In 1995, Ruth Lynden-Bell moved to Queen's University Belfast as co-founder of the interdisciplinary Atomistic Simulation Group and Professor of Condensed Matter Simulation. Between 1996 and 2003, Donald held a Senior Research Fellowship as Visiting Professor at Queen's University, Belfast. He normally went to Armagh Observatory once a week, while spending the Summers in Cambridge to work with Katz and Bi\v{c}\'{a}k. Donald and Ruth returned to Cambridge when she retired from Queen's University in 2003. Donald remained very active in research, especially in relativity, the collimation of jets, stellar dynamics and -- a new research theme -- optics.

\subsection{Exact Optics}

Donald kept a 6-inch reflector in his office. He had made the mirror himself during his undergraduate days: "Grinding, polishing, and Foucault-testing my 6 1/2 inch telescope mirror while an undergraduate taught me that persistence pays, but my lack of patience makes me poorly suited to practical work"~(\ref{LB2010}). In 1983, Donald was instrumental in urging Roderick Willstrop to explore the possibilities of optical systems that could give as wide a field as the Schmidt Telescope, but with a much shorter tube. Following E.~J.~Kibblewhite's suggestion to study 3-mirror Paul-Baker systems, Willstrop then found that a large field could be obtained by perforating the paraboloidal primary mirror, bringing the convex spherical secondary mirror nearer to the primary, and putting the tertiary concave spherical mirror behind the primary~\citep{Wi84}. When by experimenting, Willstrop had reached a field of view of $4^\circ$, Donald was still not entirely satisfied and urged him to try to reach $5^\circ$. Variants of such three mirror designs are the basis for many modern survey telescopes, including the {\it Vera Rubin Telescope}.

In 2002, Donald began a series of papers on exact optics with assistance from Willstrop. Donald knew of Karl Schwarzschild's papers of 1905, which give a complete theory of imagery in the field for any reflecting telescope with a single axis. Historically, aspheric surfaces were very expensive, so Schwarzschild concentrated attention on spherical surfaces or those for which the profiles were conic sections. Without making any such assumptions, Donald found all 2-mirror systems free of spherical aberration and coma~(\ref{LB2002},\ref{WLB}), including previously known telescope designs, such as Ritchey-Chretien or Couder. The difference between Exact Optics and all earlier papers on optical design is that Donald asked:  What shapes must two mirrors have to be free of spherical aberration and coma?  All previous opticians chose plausible mirror shapes and asked how good will the images be and what (small) changes in the shapes of the mirrors will improve the images? Donald's papers are mathematical, but they show an abiding interest in practical construction of instruments that stretches back to Donald's Meccano childhood toys.

\subsection{Last works}

Fittingly, some of Donald's last papers return to subjects associated with his Ph.~D. thesis -- both the stellar dynamics that was in it, and the magnetohydrodynamics that was meant to be in it! The proper motions of Milky Way stars supplied by the multi-epoch {\it Sloan Digital Sky Survey} data or by the {\it Gaia} satellite meant that the six components of the stellar velocity dispersion tensor in the Galaxy could finally be constructed directly from the data. Empirically, the velocity dispersion tensor is close to exact alignment in spherical polar coordinates~(\ref{ELB2016}). Inspired by this, Donald and co-workers proved the theorem that if the even part of the phase space distribution function is invariant under time reversal, then the velocity dispersion tensor must everywhere exactly aligned in ellipsoidal coordinates (which include spherical polars as a special case) and the potential must be of exactly separable or St\"ackel form.  

Donald had always been puzzled as to how swirling discs of conducting fluid around young stars, quasars and micro-quasars are able to generate highly collimated jets over distances hundreds or thousands of times the size of the disc. At first, he suspected the origin may be vortices above and below the central object that cause the beaming~(\ref{LB78c}). Later, he favoured magnetic collimation, and built a series of simplified models to gain insight. In these, magnetic forces dominate over any gas pressure in the jet. Outside the magnetic cavity of the jet, there is pressure from an ambient coronal medium. The central body and accretion disc are both massive and conducting, so they drag the magnetic field with their fluid motions. The build-up of magnetic energy due to differential motions drives the jet. Donald used the virial theorem to show that, as the upward magnetic flux of an accretion disc is twisted relative to the downward flux, the height of the magnetic field configuration grows~(\ref{LB1996},\ref{LB2006}). It assumes a vertical cylindrical geometry in which each additional twist of the field produces an equal increment in the height of the cylinder. So, after many twists, the cylinder becomes very tall and thin. Donald thought that this collimates the narrow jets seen in quasars, radio galaxies, and young stars (Herbig-Haro objects). Donald built sequences of force-free models to illustrate this, arguing that this is more insightful than magnetohydrodynamical simulations. 

\section{Influence and Legacy}

Donald passed away on 6th February 2018 at his home in Cambridge, after an earlier stroke from which he never recovered.

Donald loved mathematics and he used his skills in diverse areas of astrophysics and relativity. But, he always insisted that research must be, first and foremost, great fun. Donald worked on whatever took his fancy, from data analysis to exact optics, from thermodynamics to general relativity, from stellar dynamics to large-scale structure. He loved to work with graduate students, densely covering the blackboard with chalked equations; he loved to ask questions in seminars with his booming voice, often from an unusual perspective; he loved to talk astronomy over breakfast ... coffee, lunch or dinner ...  explaining his new ideas to sometimes bewildered listeners; he loved to lecture in his buoyant and chaotic style, especially about his own research. He bubbled over with indefatigable enthusiasm for any topic in astronomy or related fields. He had something of the boisterousness, energy and exuberance of A.A. Milne's Tigger.

Donald was at the pinnacle of his creative powers in the 1960s and early 1970s. Amongst the welter of fine papers, three stand out as enduring achievements of the very highest calibre -- the discovery of swing amplification in gas discs with Goldreich; the {\it Nature} paper that won him the Kavli Prize for the suggestion that dead quasars reside in the centres of galaxies; and the paper with Pringle that elucidated the workings of accretion discs. Further, the accumulated papers on stellar dynamics from his 1960 Ph.~D thesis to his 1973 Saas-Fee Lectures form a hugely impressive corpus of work. This was the first sustained and vigorous attack on the problems of the structure and dynamics of galaxies by a powerful thinker since the days of Eddington and Jeans -- almost exclusively using analytic methods. Galactic dynamics became more computational from the 1980s onwards with numerical simulations playing an increasingly important role. This was not quite to Donald's taste, and so his imagination took him on to other astronomical problems -- relativity, Mach's principle, exact optics -- where he could use pen, paper and mathematical prowess.

Donald's achievements were recognized by a succession of medals and awards from both national and international learned societies, but one that gave him great pleasure was his election to the Presidency of the Royal Astronomical Society (RAS). He regarded the RAS as playing a crucial role in UK astronomy and it was a very important part of his life. Donald was a tremendous mentor and colleague. He supervised over fifty students, including Simon White FRS 1997 (\lq\lq Simon just came and told me interesting things"), Wyn Evans, Ofer Lahav, Michael Penston, Jim Collett, Mike Hudson, Christophe Pichon and Somak Raychaudhury. Though not formally supervisor, Donald also played an important role in the Ph.~D. work of Ken Freeman FRS 1998, Jim Pringle and Tim de Zeeuw. 

Donald did not enjoy administration, but he was a very conscientious individual. He took over the Directorship of the Institute of Astronomy at a perilous moment, yet thanks to his stewardship -- and that of Martin Rees -- the Institute has thrived over the following decades. Elsewhere, there are personal reminiscences of Donald from his friends and colleagues~\citep{Rees18,Ev18,La18}. Here, it suffices to say that very few people will carve so distinctive and memorable a trail in astronomy -- and in life -- the way that Donald Lynden-Bell did.

\section*{Acknowledgments}
The author is very grateful to Ruth Lynden-Bell for advice, help and comments, as well as access to Donald's papers and private writings. She also made available her personal collection of photographs of Donald. The author is indebted to many people for useful criticism of the text, including Jiri Bi\v{c}\'{a}k, Jim Collett, Douglas Gough, Elizabeth Griffin, Mark Hurn, Ofer Lahav, Malcolm Longair, Jim Pringle, Martin Rees, Alison Rose, Amanda Smith, Scott Tremaine, Alar Toomre, Terry Wall and Roderick Willstrop.

\section*{AWARDS AND RECOGNITION}
\begin{itemize}\itemsep=0pt
\item[]Balzan Prize (1983)
\item[]RAS Eddington Medal (1984)
\item[]AAS Brouwer Award (1990)
\item[]Membership of National Academy of Sciences (1990)
\item[]Oort Visiting Professorship, Leiden (1992)
\item[]RAS Gold Medal (1993)
\item[]Bruce Medal of Astronomical Society of the Pacific (1998)
\item[]Henry Norris Russell Lectureship, AAS (2000)
\item[]Commander of the British Empire (2000)
\item[]John J Carty Award for the Advancement of Science (2000)
\item[]Blauuw Visiting Professorship, Groningen (2007)
\item[]Kavli Prize for Astrophysics (2008)
\end{itemize}

\newpage

\section*{Brief Author Profile}
\bigskip

{\bf Wyn Evans} is Professor of Astrophysics at the University of Cambridge. He was a graduate student of Donald Lynden-Bell from 1985-1988. He subsequently was a Junior and Senior Research Fellow of King's College, Cambridge and a Lindemann Fellow at the Massachusetts Institute of Technology. He was Reader in Theoretical Physics, Oxford University, before returning to Cambridge in 2003. His research interests include galactic structure and dynamics, the dark matter problem, astroparticle physics, Survey science, solar system dynamics and gravitational lensing.

\newpage

\section*{References}

\begin{donaldpapers}

 \Dbibitem{LB1962a}(1962)
 Stellar dynamics. Potentials with isolating integrals,
  \emph{Monthly Notices of the Royal Astronomical Society} {\bf 124} 95-123.
  (doi: \doi{10.1093/mnras/124.2.95})

\Dbibitem{LB1962b}(1962)
 Stellar dynamics. Exact Solution of the Self-gravitation Equation,
  \emph{Monthly Notices of the Royal Astronomical Society} {\bf 123} 447-458.
  (doi: \doi{10.1093/mnras/123.5.447})
 
\Dbibitem{ELS}(1962)
 (With O.~Eggen and A.~R.~Sandage) Evidence from the motions of old stars that the Galaxy collapsed.
  \emph{The Astrophysical Journal} {\bf 136} 748-766.
  (doi: \doi{10.1086/147433})
 
\Dbibitem{LBecc}(1963)
The invariant eccentricity of galactic orbits
\emph{The Observatory},
{\bf 83} 23-25

\Dbibitem{GLB1}(1965)
(With P.~Goldreich) Gravitational stability of uniformly rotating disks.
\emph{Monthly Notices of the Royal Astronomical Society} {\bf 130} 97-124.
 (doi: \doi{10.1093/mnras/130.2.97})
          
\Dbibitem{GLB2}(1965)
(With P.~Goldreich) Spiral arms as sheared gravitational instabilities.
 \emph{Monthly Notices of the Royal Astronomical Society} {\bf 130} 125-158.
(doi: \doi{10.1093/mnras/130.2.125})

\Dbibitem{LB1967Mach}(1967)
On the origins of space-time and inertia.
\emph{Monthly Notices of the Royal Astronomical Society}
{\bf 135} 413-428.
(doi: \doi{10.1093/mnras/135.4.413})
       
\Dbibitem{LB1967}(1967)
Statistical mechanics of violent relaxation in stellar systems.
\emph{Monthly Notices of the Royal Astronomical Society}
{\bf 136} 101-121.
(doi: \doi{10.1093/mnras/136.1.101})

\Dbibitem{LBCoop}(1967)
Cooperative Phenomena in Stellar Dynamics.
In {\it Relativity Theory and Astrophysics. Vol.2: Galactic Structure}, Lectures in Applied Mathematics, Vol. 9. Edited by J{\"u}rgen Ehlers.  Providence, Rhode Island: American Mathematical Society. 

\Dbibitem{LBO}(1967)
(With J.~P.~Ostriker)
On the stability of differentially rotating bodies.
\emph{Monthly Notices of the Royal Astronomical Society}
{\bf 136} 293-310.
(doi: \doi{10.1093/mnras/136.3.293})

\Dbibitem{LB1968}(1968)
(With R.~Wood)
The gravo-thermal catastrophe in isothermal spheres and the onset of red-giant structure for stellar systems.
\emph{Monthly Notices of the Royal Astronomical Society}
{\bf 138} 495-525.
(doi: \doi{10.1093/mnras/138.4.495})

\Dbibitem{GLB3}(1969)
(With P.~Goldreich)
Io, a jovian unipolar inductor.
\emph{The Astrophysical Journal}
{\bf 156} 59-78.
(doi: \doi{10.1086/149947}

\Dbibitem{LBS}(1969)
(With N.~Sanitt)
The Schr{\"o}dinger operator criterion for the stability of galaxies and gas spheres.
\emph{Monthly Notices of the Royal Astronomical Society}
{\bf 143} 167-187.
(doi: \doi{10.1093/mnras/143.2.167})

\Dbibitem{LB1969}(1969)
Galactic Nuclei as Collapsed Old Quasars.
\emph{Nature} {\bf 223} 690-694
(doi: \doi{10.1038/223690a0})

\Dbibitem{LBR}(1971)
(With M.~J.~Rees)
On quasars, dust and the galactic centre.
\emph{Monthly Notices of the Royal Astronomical Society}
{\bf 152} 461-474.
(doi: \doi{10.1093/mnras/152.4.461})

\Dbibitem{LBK1972}(1972)
(With A.~J. Kalnajs)
On the generating mechanism of spiral structure
\emph{Monthly Notices of the Royal Astronomical Society}
{\bf 157}, 1-30
(doi: \doi{10.1093/mnras/157.1.1})

\Dbibitem{LB73}(1973)
Topics in the Dynamics of Stellar Systems
In {\it Saas-Fee Advanced Course: Dynamical Structure and Evolution of Stellar Systems}, eds. L. Martinet, M. Mayor, Geneva Observatory.

\Dbibitem{LBP74}(1974)
(With J.~E.~Pringle)
The evolution of viscous discs and the origin of the nebular variables.
\emph{Monthly Notices of the Royal Astronomical Society}
{\bf 168} 603-637.
(doi: \doi{10.1093/mnras/168.3.603})

\Dbibitem{LB1976}(1976)
Dwarf galaxies and globular clusters in high velocity hydrogen streams.
\emph{Monthly Notices of the Royal Astronomical Society}  
{\bf 174} 695-710.
(doi: \doi{10.1093/mnras/174.3.695})

\Dbibitem{LL77}(1977)
(With D.~N.~C.~Lin)
Simulation of the Magellanic Stream to estimate the total mass of the Milky Way.
\emph{Monthly Notices of the Royal Astronomical Society}    
{\bf 181} 59-81.
(doi: \doi{10.1093/mnras/181.2.59})

\Dbibitem{LB1977}(1977)
(With R.~M.~Lynden-Bell) On the negative specific heat paradox.
\emph{Monthly Notices of the Royal Astronomical Society}
{\bf 181} 405-419.
(doi: \doi{10.1093/mnras/181.3.405})

\Dbibitem{LB1978a}(1978)
(With S.~Pineault)
Relativistic disks - 1. Counter rotating disks.
\emph{Monthly Notices of the Royal Astronomical Society}
{\bf 185} 679-694.
(doi: \doi{10.1093/mnras/185.4.679})

\Dbibitem{LB1978b}(1978)
(With S.~Pineault)
Relativistic disks - 11. Self-similar disks in rotation.
\emph{Monthly Notices of the Royal Astronomical Society}
{\bf 185} 695-712.
(doi: \doi{10.1093/mnras/185.4.695})

\Dbibitem{LB78c}(1978)
Gravity power.
\emph{Physics Scripta}
{\bf 17} 185-191.
(doi: \doi{10.1088/0031-8949/17/3/009})

\Dbibitem{LB79bar}(1979)
On a mechanism that structures galaxies.
\emph{Monthly Notices of the Royal Astronomical Society}
{\bf 187} 101-107.
(doi: \doi{10.1093/mnras/187.1.101})

\Dbibitem{LB1980}(1980)
(With P.~P.~Eggleton) 
On the consequences of the gravothermal catastrophe.
\emph{Monthly Notices of the Royal Astronomical Society}
{\bf 191} 483-498.
(doi: \doi{10.1093/mnras/191.3.483})

\Dbibitem{LL82}(1982)
(With D.~N.~C.~Lin)
On the proper motion of the Magellanic Clouds and the halo mass of our galaxy.
\emph{Monthly Notices of the Royal Astronomical Society}
{\bf 198} 707-721.
(doi: \doi{10.1093/mnras/198.3.707})

\Dbibitem{LB1982}(1982)
The Fornax-Leo-Sculptor system.
\emph{The Observatory} {\bf 102} 202-208.

\Dbibitem{LB1983}(1983)
The origin of dwarf spheroidal galaxies.
In {\it Internal Kinematics and Dynamics of Galaxies}, IAU Symposium 100, 89-91. Edited by E. Athanassoula, Reidel, Dordrecht.

\Dbibitem{ZL85}(1985)
 (With P.~T.~de Zeeuw)
 Best approximate quadratic integrals in stellar dynamics.
\emph{Monthly Notices of the Royal Astronomical Society} 
{\bf 215} 713-730.
(doi: \doi{10.1093/mnras/215.4.713})
  
\Dbibitem{THL}(1986)
(With S.~Tremaine and M. H\'enon)
H-functions and mixing in violent relaxation
\emph{Monthly Notices of the Royal Astronomical Society}
{\bf 219} 285-297.
(doi: \doi{10.1093/mnras/219.2.285})

 \Dbibitem{Dr87}(1987)
(With S.~M.~Faber, D.~Burstein, R.~L.~Davies, A.~Dressler, R.~J.~Terlevich and G.~Wegner)
\emph{The Astrophysical Journal}
Spectroscopy and Photometry of Elliptical Galaxies. I. New Distance Estimator
\emph{The Astrophysical Journal}
{\bf 313} 42-58.
(doi: \doi{10.1086/164947})

\Dbibitem{LB88}(1988)
(With S.~M.~Faber, D.~Burstein, R.~L.~Davies, A.~Dressler, R.~J.~Terlevich and G.~Wegner)
Photometry and Spectroscopy of Elliptical Galaxies. V. Galaxy Streaming toward the New Supergalactic Center
\emph{The Astrophysical Journal}
{\bf 326} 19-49.
(doi: \doi{10.1086/166066})

\Dbibitem{LLR}(1988)
(With O.~Lahav and M.~Rowan-Robinson)
The peculiar acceleration of the Local Group as deduced from the optical and IRAS flux dipoles
\emph{Monthly Notices of the Royal Astronomical Society}
{\bf 234} 677-701
(doi: \doi{10.1093/mnras/234.3.677})

\Dbibitem{LL88}(1988)
(With J.~P.~S.~Lemos)
On Penston's self-similar solution for cold collapse.
\emph{Monthly Notices of the Royal Astronomical Society}
{\bf 233} 197-208.
(doi: \doi{10.1093/mnras/233.1.197})

\Dbibitem{LL89}(1989)
(With J.~P.~S.~Lemos)
A General Class of Spherical Newtonian Self-Similar Solutions for a Cold Fluid - Part Two - Solutions with Gravity.
\emph{Monthly Notices of the Royal Astronomical Society}
{\bf 240} 317-327
(doi: \doi{10.1093/mnras/240.2.317})

\Dbibitem{LLB89}(1989)
(With O.~Lahav and D.~Burstein)
Cosmological deductions from the alignment of local gravity and motion.
\emph{Monthly Notices of the Royal Astronomical Society}
{\bf 241} 325-345
(doi \doi{10.1093/mnras/241.2.325})

\Dbibitem{Ev90}(1990)
 (With N.~W.~Evans and P.~T.~de Zeeuw)
 The flattened isochrone.
 \emph{Monthly Notices of the Royal Astronomical Society}  
 {\bf 244} 111-129.

\Dbibitem{BLKdisc}(1993)
(With J.~Katz and J.~Bi\v{c}\'{a}k)
Relativistic disks as sources of static vacuum spacetimes.
\emph{Physical Review D}
{\bf 47} 4334-4343.
(doi: \doi{10.1103/PhysRevD.47.4334})

\Dbibitem{BLP}(1993)
(With J.~Bi\v{c}\'{a}k and C.~Pichon)
Relativistic Discs and Flat Galaxy Models.
\emph{Monthly Notices of the Royal Astronomical Society}
{\bf 265} 126-143.
(doi: \doi{10.1093/mnras/265.1.126})

\Dbibitem{LBDwin}(1994)
(With R.~C.~Kraan-Korteweg, A.~J.~Loan, W.~B.~Burton, O.~Lahav, H.~C.~Ferguson and P.~A.~Henning)
Discovery of a nearby spiral galaxy behind the Milky Way
\emph{Nature}
{\bf 372} 77-79
(doi: \doi{10.1038/372077a0})

\Dbibitem{LBLB1995}(1995)
(With R.~M.~Lynden-Bell)
Ghostly streams from the formation of the Galaxy's halo.
\emph{Monthly Notices of the Royal Astronomical Society}  
{\bf 275} 429-442.
(doi: \doi{10.1093/mnras/275.2.429})
   
\Dbibitem{LB1995KB}(1995)
(With J.~Katz and J.~Bi\v{c}\'{a}k)
Mach's principle from the relativistic constraint equations.
\emph{Monthly Notices of the Royal Astronomical Society}
{\bf 272} 150-160.
(doi: \doi{10.1093/mnras/272.1.150})

\Dbibitem{FLB}(1995)
(With K.~Fisher, O.~Lahav, Y.~Hoffman and S.~Zaroubi)
        Wiener reconstruction of density, velocity and potential fields from all-sky galaxy redshift surveys.
\emph{Monthly Notices of the Royal Astronomical Society}        
{\bf 272} 885-908
(doi: \doi{10.1093/mnras/272.4.885})

\Dbibitem{LB1996}(1996)
Magnetic collimation by accretion discs of quasars and stars.
\emph{Monthly Notices of the Royal Astronomical Society}
{\bf 279} 389-401.
(doi: \doi{10.1093/mnras/279.2.389})

\Dbibitem{KBL}(1997)
(With J.~Katz and J.~Bi\v{c}\'{a}k)
Relativistic conservation laws and integral constraints for large cosmological perturbations.
\emph{Physical Review D}
{\bf 55} 5957-5969.
(doi: \doi{10.1103/PhysRevD.55.5957})

\Dbibitem{LB1999}(1999)
Negative Specific Heat in Astronomy, Physics and Chemistry.
\emph{Physica A: Statistical Mechanics and its Applications}
{\bf 263} 293-304.
(doi: \doi{10.1016/S0378-4371(98)00518-4})

\Dbibitem{LBObitHoyle}(2001)
Obituary: Sir Fred Hoyle (1915-2001).
\emph{The Observatory}, {\bf 121} 405-408

\Dbibitem{LB2002}(2002)
Exact optics: a unification of optical telescope design.
\emph{Monthly Notices of the Royal Astronomical Society}
{\bf 334} 787-796.
(doi: \doi{10.1046/j.1365-8711.2002.05486.x})

\Dbibitem{WLB}(2003)
(With R.~V.~Willstrop)
Exact optics - II. Exploration of designs on- and off-axis.
\emph{Monthly Notices of the Royal Astronomical Society}
{\bf 342} 33-49.
(doi: \doi{10.1046/j.1365-8711.2003.06434.x})
   
\Dbibitem{LB2003}(2003)
On why discs generate magnetic towers and collimate jets.
\emph{Monthly Notices of the Royal Astronomical Society}
{\bf 341} 1360-1372.
(doi: \doi{10.1046/j.1365-8711.2003.06506.x})

\Dbibitem{LB2006}(2006)
Magnetic jets from swirling discs
\emph{Monthly Notices of the Royal Astronomical Society}
{\bf 369} 1167-1188
(doi = \doi{10.1111/j.1365-2966.2006.10349.x})

\Dbibitem{BKL}(2007)
(With J.~Katz and J.~Bi\v{c}\'{a}k)
Cosmological perturbation theory, instantaneous gauges, and local inertial frames.
\emph{Physical Review D} 
{\bf 76} 063501-063532
(doi: \doi{10.1103/physrevd.76.063501})
 
\Dbibitem{Be07}(2007)
(With V.~A.~Belokurov, N.~W.~Evans, M.~J.~Irwin and others)  
An Orphan in the ``Field of Streams''.
\emph{The Astrophysical Journal}
{\bf 658} 337-344.
(doi: \doi{10.1086/511302})

\Dbibitem{JL}(2007)
(With S.~Jin)
Are Complex A and the Orphan stream related?
\emph{Monthly Notices of the Royal Astronomical Society}
{\bf 378} L64-L66.
(doi: \doi{10.1111/j.1745-3933.2007.00321.x})

\Dbibitem{LB2010}(2010)
 Searching for Insight.
\emph{Annual Reviews of Astronomy and Astrophysics}   
{\bf 48} {1-19}.
(doi: \doi{10.1146/annurev-astro-081309-130859})
       
\Dbibitem{ELB2016}(2016)
(With N.~W.~Evans, J.~L.~Sanders, A.~A.~Williams, J.~An and W.~Dehnen)
The alignment of the second velocity moment tensor in galaxies.
\emph{Monthly Notices of the Royal Astronomical Society}
{\bf 456} 4506-4523.
(doi: \doi{10.1093/mnras/stv2729})

\Dbibitem{LB2016}(2016)
Principal velocity surfaces in stellar dynamics.
\emph{Monthly Notices of the Royal Astronomical Society}
{\bf 458} 726-732
(doi: \doi{10.1093/mnras/stw229})

\end{donaldpapers}

\renewcommand{\refname}{References to Other Authors}
\bibliographystyle{QCP_Master_doi}
\bibliography{Non-Donald_Papers}

%

\end{document}